\documentclass[useAMS,usenatbib,usegraphicx]{mn2e}

\newcommand{\etal}{{\rm et al.\ }}
\bibpunct{(}{)}{;}{a}{}{,}

\title[The Spatial Variation of the 3$\mu$m Dust Features in Circinus]{The Spatial Variation of the 3$\mu$m Dust Features in Circinus}
\author[M. D. Colling, P. F. Roche and R. E. Mason]{M. D. Colling$^{1}$\thanks{E-mail:
mdc@astro.ox.ac.uk}, P. F.
Roche$^{1}$ and R. E. Mason$^{2}$\\
$^{1}$Astrophysics, Department of Physics, University of Oxford, DWB, Keble Road, Oxford OX1 3RH\\
$^{2}$Gemini Observatory, Northern Operations Center, Hilo, HI 96720}
\begin{document}

\date{Accepted 0000 December 00. Received 0000 December 00; in original form 0000 November 00}

\pagerange{\pageref{firstpage}--\pageref{lastpage}} \pubyear{2008}

\maketitle

\label{firstpage}

\begin{abstract}
We report spatially-resolved variations in the 3.4-$\mu$m hydrocarbon absorption feature and the 3.3-$\mu$m polycyclic aromatic hydrocarbon (PAH) emission band in the Circinus galaxy over the central few arcsec. The absorption is measured towards warm emitting dust associated with Coronal line regions to the east and west of the nucleus.  There is an absorption optical depth $\tau_{{3.4\mu}m}\sim$0.1 in the core which decreases to the west and increases to the east. This is consistent with increased extinction out to $\sim$40 pc east of the core, supported by the Coronal emission line intensities which are significantly lower to the east than the west. PAH emission is measured to be symmetrically distributed out to $\pm$4 arcsec, outside the differential extinction region. The asymmetry in the 3.4-$\mu$m absorption band reflects that seen in the 9.7-$\mu$m silicate absorption band reported by \citet{Roche06} and the ratio of the two absorption depths remains approximately constant across the central regions, with $\tau_{{3.4\mu}m}$/$\tau_{{9.7\mu}m}\sim$0.06$\pm$0.01. This indicates well-mixed hydrocarbon and silicate dust populations, with no evidence for significant changes near the nucleus.
\end{abstract}


\section{Introduction}

\begin{figure}
\begin{center}
\includegraphics[width=8cm]{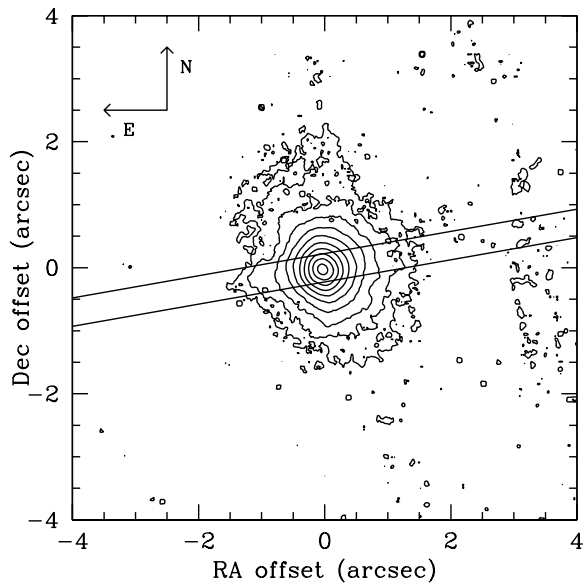}
\caption{The GNIRS slit superimposed on the NACO $L$-band image of the Circinus nucleus, taken from the ESO Science Archive Facility and previously published in \citet{Prieto04}. The image is centred at the $L$-band emission peak, which is assumed to coincide with the active nucleus. The contours are logarithmic and separated by a factor of log(2).
} 
\label{fig:contour}
\end{center}
\end{figure}

\begin{figure}
\begin{center}
\includegraphics[width=8.8cm]{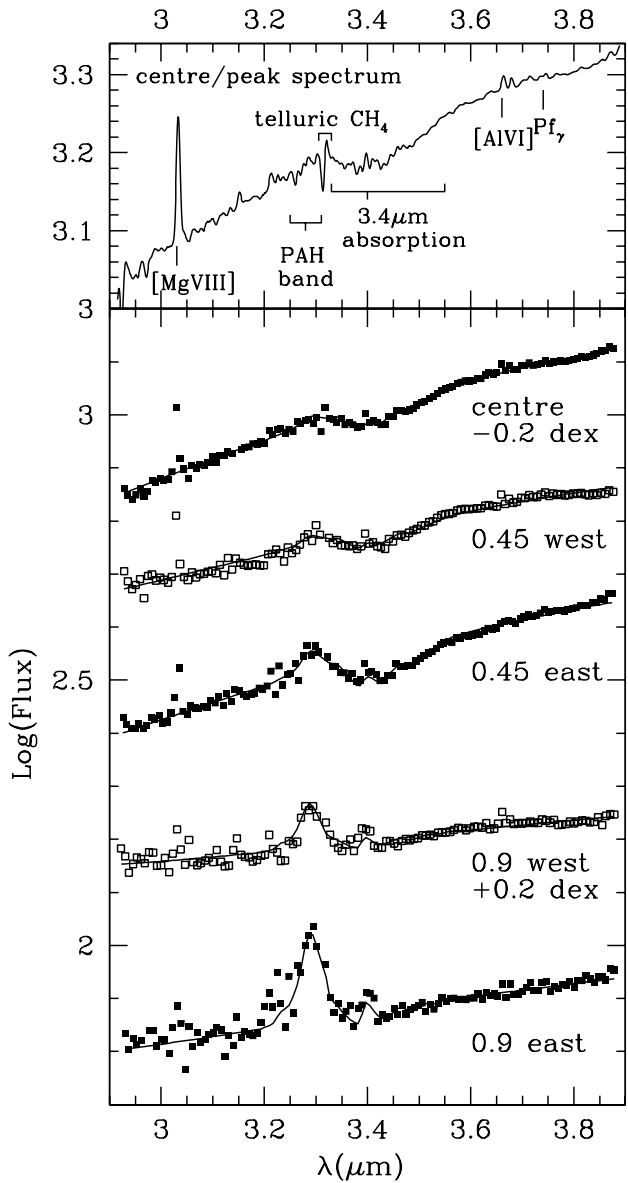}
\caption{Five of the spectra used to measure the spatial variation of the 3~-~4~$\mu$m absorption and emission. The centre spectrum together with the east and west spectra at 0.45 and 0.9 arcsec intervals. The lines are the 4-component model fits described in section \ref{sec:model}. The pixels have been binned on the dispersion axis for plotting and fitting and the fluxes are offset by the values stated as appropriate. The flux is in units of $10^{-20}$ W cm$^{-2}$ $\mu{\rm{m}}^{-1}$. The top panel shows the peak spectrum at full resolution and smoothed, with the lower panel showing the spectra after binning and smoothing.
} 
\label{fig:spectrum}
\end{center}
\end{figure}

The Circinus galaxy is the nearest Active Galactic Nucleus (AGN) to the Milky Way at $\sim$4 Mpc distance. The galaxy is a SAb spiral inclined at about 65$^{\circ}$, suffering Galactic extinction of ${A_v}\sim$1.7 mag (\citealt{Freeman77}; \citealt{Jones99}). The nucleus is a Seyfert 2, with high-excitation emission lines and an obscured broad line emission region revealed by polarimetry \citep{Oliva99}. A strong ionization cone is detected in the optical, particularly well defined by HST [{O\,\sc iii}] filter imaging in \citet{Wilson00}. A star formation ring is identified about the Circinus galaxy nucleus at a radius of $\sim$10 arcsec from H$_\alpha$ and [{S\,\sc ii}] imaging \citep{Marconi94}.

Because of the proximity of the Circinus galaxy, hereafter Circinus, spectroscopy at high spatial resolution can be undertaken, with 1 arcsec corresponding to $\sim$20 pc. Observations of Circinus in the mid-infra-red (MIR) at 10 $\mu$m have showed spatial variations in the silicate-dust absorption spectra within $\pm$2 arcsec from the nucleus. The variation is asymmetric and consistent with there being an extended dusty torus-like structure of diameter $\sim$80 pc around the Seyfert 2 core, inclined at an angle similar to the disc of the galaxy \citep{Roche06}. Interferometric observations in the MIR by \citet{Tristram07} are well fit with a clumpy torus model for a cool dust component extending out to 1-pc, half-light radius. Whether these components are part of a continuous structure, and whether they are further linked to a water maser disc of diameter 0.2-0.8 pc around the core, aligned with the inner and outer dust structures and nearly orthogonal to the ionization cone \citep{Greenhill03}, is unknown.

Studies of the 3.4-$\mu$m hydrocarbon absorption in the Milky Way have shown that it arises from the diffuse interstellar medium (ISM) (e.g. \citealt{Pendleton94}; \citealt{Chiar02}), the optical depth varying greatly with position and closely correlated with the visual extinction. The absorption is deepest toward the Galactic centre and weaker toward stellar regions, suggesting that the aliphatic hydrocarbons responsible are not uniformly distributed in the Galaxy, but concentrated toward the centre \citep{Sandford95}.  The thermal background due to sky emission in the $L$-band makes spatially-resolved observations of the absorption feature difficult to make from the ground, as the flux and hence the signal-to-noise ratio decreases greatly off the core.  A VLT/NACO image of Circinus in the $L$-band shows a very compact nucleus, of 1-2 arcsec diameter, with a core FWHM of 0.185 arcsec \citep{Prieto04}, meaning that high spatial resolution and relatively long integration times are required to investigate any spatial structure.

Polycyclic Aromatic Hydrocarbons (PAHs) are identified as the emitting molecules of distinctive infra-red bands beyond 3-$\mu$m (\citealt{Leger84}; \citealt{Allamandola85}). PAHs represent the smallest grain sizes of the dust population, excited by the far-UV (\citealt{Sellgren84}; \textit{see review in} \citealt{Tielens08}) and destroyed by hard-UV fluxes (\citealt{Aitken85}; \citealt{Leach89}). PAH emission bands were observed to be prominent in the large aperture (14x20 arcsec) ISO spectra of Circinus \citep{Moorwood96}, but absent in a small aperture (4.2 arcsec diameter), where the flux is dominated by the nuclear emission (\citealt{Roche91}). Observations of the 11.3-$\mu$m band in Circinus revealed PAH emission to be unaffected by the differential extinction detected in the nucleus, and to have a very different spatial distribution from the MIR dust continuum emission \citep{Roche06}. Though PAH band ratios can vary, the 3.3 and 11.3 $\mu$m features are expected to have similar spatial distributions as both are thought to arise from neutral PAH emission \citep{Allamandola99}.

The primary object of this study is to present long-slit spectroscopic observations of the 3.4-$\mu$m hydrocarbon dust absorption feature and the 3.3-$\mu$m emission band, and compare their spatial distributions to those of the silicate absorption and MIR dust emission components.

\section{Data}

Spectra were taken in March 2007, programme GS-2007A-Q-41, at the Gemini South telescope with the Gemini Near-Infrared Spectrograph (GNIRS) \citep{Elias06}. The instrument configuration covered a range of 2.9~-~3.9 $\mu$m with a 0.45 arcsec wide slit with 32/mm grating, $R = \frac{\lambda}{{\Delta}\lambda} = 1700$. The red sensitive, short focal length camera was used with a scale of 0.15 arcsec pixel$^{-1}$. The slit alignment, east/west PA 100$^{\circ}$ (see Fig. \ref{fig:contour}), was chosen to coincide with the extended emission observed in the MIR and the spatially extended spectra in that band from \citet{Roche06}.

The telescope was nodded along the slit in 10 arcsec steps. This means that the extracted spectra are effectively differential measurements compared to reference positions 10 arcsec from the nucleus.  However, here we investigate the circumnuclear emission on scales $<$ 5 arcsec from the core, and so the effects of emission in the reference nod positions are small. The total integration time for Circinus was $\sim$48~min, while for the telluric standard star, HIP 69972, it was $\sim$8~min. The FWHM of the PSF is estimated to be 0.5~arcsec, at 3.79-$\mu$m, from the observation of the comparison star. 

The reduction was undertaken with {\sc IRAF} v2.12.2a and Gemini {\sc IRAF} v1.9 using the GNIRS reduction package \citep{Cooke05}. The spectra were flat-fielded, the frames coadded and a wavelength calibration applied based on the sky-emission lines that are prominent in the near-IR region. There was some evidence of detector cross-talk and a systematic, but low-level, diagonal artefact. Both of these were partially corrected for by subtracting backgrounds, composed of an average of unexposed regions, from every line of the exposed regions of the spectra. One dimensional extractions of the spectra were then produced. These were telluric-corrected and flux-calibrated, using the extracted spectrum of the standard, individually to minimise noise. The 1D spectra were binned in 0.45 arcsec intervals, comparable to the FWHM and to increase signal.

Continuum fits to the spatially-binned spectra were performed using Legendre polynomials, of order 2 or 3 dependent on the curve of the fitted spectrum. The continuum regions fit were the clear regions at 29600:29980, 31680:31860, 35940:36200, 38230:38420 ${\rm{\AA}}$ with particular care taken to ensure a reasonable fit across the 3.17~-~3.6 $\mu$m region. Line fluxes were found by fitting Gaussians to the lines and integrating the flux. The PAH flux and 3.4-$\mu$m feature optical depth were found using the continuum fits, from which it was decided to separate those components from the continuum flux using model fits to the spectra, described in section \ref{sec:model}. Examples of the final spectra, displayed with the model fits described below, are shown in Fig. \ref{fig:spectrum}.

\section{Results}

\begin{figure}
\begin{center}
\includegraphics[width=8.4cm]{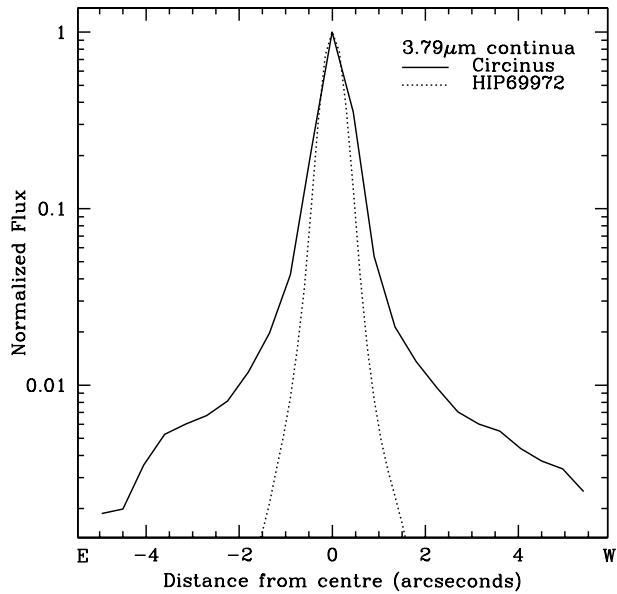}
\caption{The normalized spatial profiles of the continuum at 3.79-$\mu$m in Circinus and the reference star. 
}
\label{fig:fluxwing}
\end{center}
\end{figure}

\begin{figure*}
\begin{center}
\includegraphics[width=16.8cm]{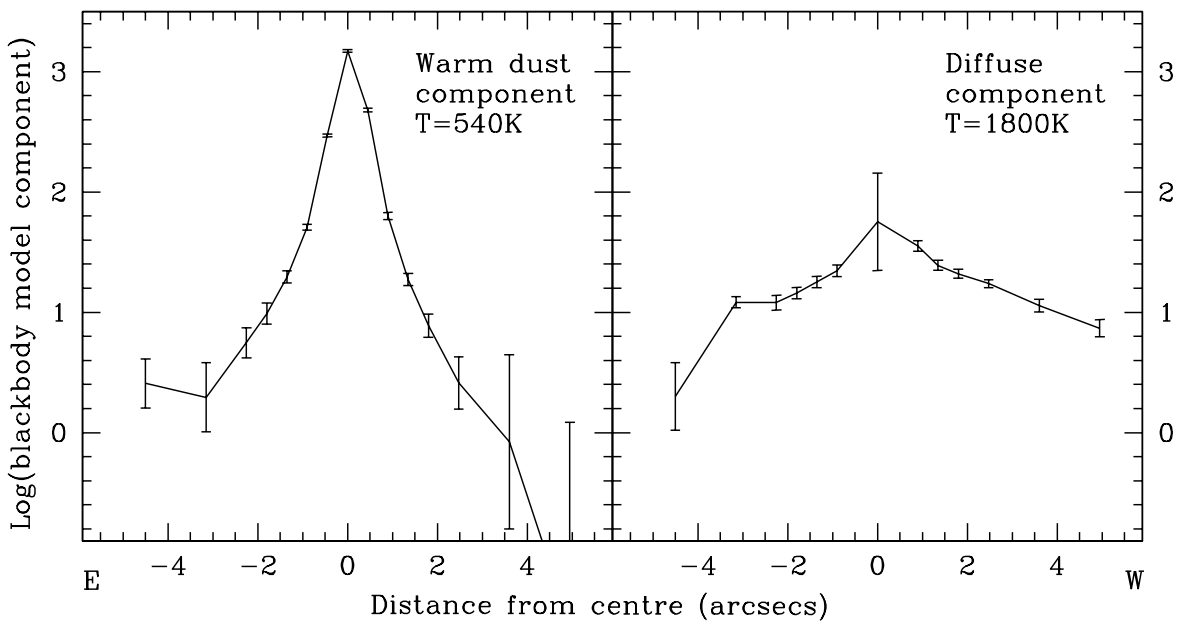}
\caption{The spatial distributions of the continuum components from the model fits, described in section \ref{sec:model}. The warm dust component is fixed at 540 K and represents the dust emission from the core. The hot component is fixed at 1800 K and represents diffuse photospheric and dust emission from stars in the host galaxy. The warm dust component is found to peak centrally very strongly and then decrease rapidly further from the core. The stellar component is also peaked centrally, but is proportionally weaker, and decreases more slowly with distance from the core. The combination of these two components reflects the shape of the continuum dust emission observed at 3.79-$\mu$m in Fig. \ref{fig:fluxwing}.}
\label{fig:model_warmcold}
\end{center}
\end{figure*}

The extracted spectra show clearly the 3.4-$\mu$m absorption, the 3.3-$\mu$m emission and several spectral emission lines on top of the continuum. The compact nature of the core in Circinus is evident, as expected from the \citet{Prieto04} imaging. Fig. \ref{fig:fluxwing} shows the emission is a combination of a compact nuclear component together with a more extended component attributed to diffuse galactic emission. The diffuse emission starts to dominate beyond $\pm$2 arcsec as seen by the change in gradient at these positions.

Model fits to the spatially-extracted, binned spectra have been used to find the absolute and relative values of the primary components in the spectra. Section \ref{sec:model} describes the model and the following sections analyse the 3.4-$\mu$m feature and the PAH emission. The spectra were analysed from the peak flux outwards, with the eastward and westward sides being compared to look for evidence of asymmetry. Spatially-extended emission from hydrogen and forbidden emission lines have also been measured and analysed and this is reported in the final section.

\subsection{Model fits}
\label{sec:model}

A model consisting of four components has been fitted to the data. A warm blackbody with T=540 K represents the dust emission from the inner core. The 3.4-$\mu$m absorption feature was applied to this component. The spectral profile used was that of the spectrum of IRS-6E in the Galactic centre (GC) from \citet{Pendleton94}. This shows the 3.4-$\mu$m hydrocarbon feature with well-defined double peaks in the optical depth at 3.38 and 3.42 $\mu$m. A second profile, from \citet{Adamson03} of the GC was also tested. This shows an extended bimodal absorption profile from 3.2 to 3.6 $\mu$m, with features from the 3.4-$\mu$m aliphatic hydrocarbon band described above and a broader 3.3-$\mu$m absorption, attributed to PAHs. However, this introduced a degeneracy between the absorption and PAH emission that cannot be resolved. The IRS-6E profile was used to produce the final fits.

The other two components were the PAH emission complex at 3.2~-~3.6~$\mu$m, obtained from the spectrum of  planetary nebula NGC 7027 \citep{Roche96}, and a hot blackbody component, fixed at 1800~K.  The PAH component includes the prominent peak at 3.28-$\mu$m and the weaker peaks and plateau emission between 3.4 and 3.6~$\mu$m.  The profile was extracted by fitting a linear continuum to the flux at 3.1 and 3.7-$\mu$m, and provides an adequate match to the PAH emission in Circinus. The hot component represents diffuse galactic emission from stellar photospheres, hot dust emission from circumstellar shells and the diffuse dust continuum, which together combine to give an effective temperature of $\sim$1800~K in this spectral region. It is expected to become increasingly important in the spectra further from the core, as the diffuse galaxy starlight and dust emission begins to dominate the continuum. Initially the temperatures of the two blackbodies were allowed to float and were found to roughly converge on the temperatures above. These were then fixed to decrease any degeneracy of components when converging the fits with the PAH and 3.4-$\mu$m components. The models were applied using a chi-squared convergence algorithm, to find the best fit to the data for the relative strengths of the four components.

Fig. \ref{fig:model_warmcold} shows the contributions to the continuum from the warm dust population and the diffuse hot component. Both populations are symmetric and centrally peaked, with the warm dust contributing an order of magnitude more flux on the nucleus. The diffuse emission proportion increases with distance from the core. At the central position, the contribution from the diffuse T=1800~K component is small compared to the warm component and the uncertainties very large. This distribution justifies the application of the absorption profile to only the warm dust component in the model. The separation of the stellar emission from the absorption also has the benefit of removing any quenching of the absorption feature as suggested in \citet{Imanishi00}.

\subsection{3.4$\mu$m hydrocarbon absorption}
\label{sec:3.4}

\begin{figure*}
\begin{center}
\includegraphics[width=8.4cm]{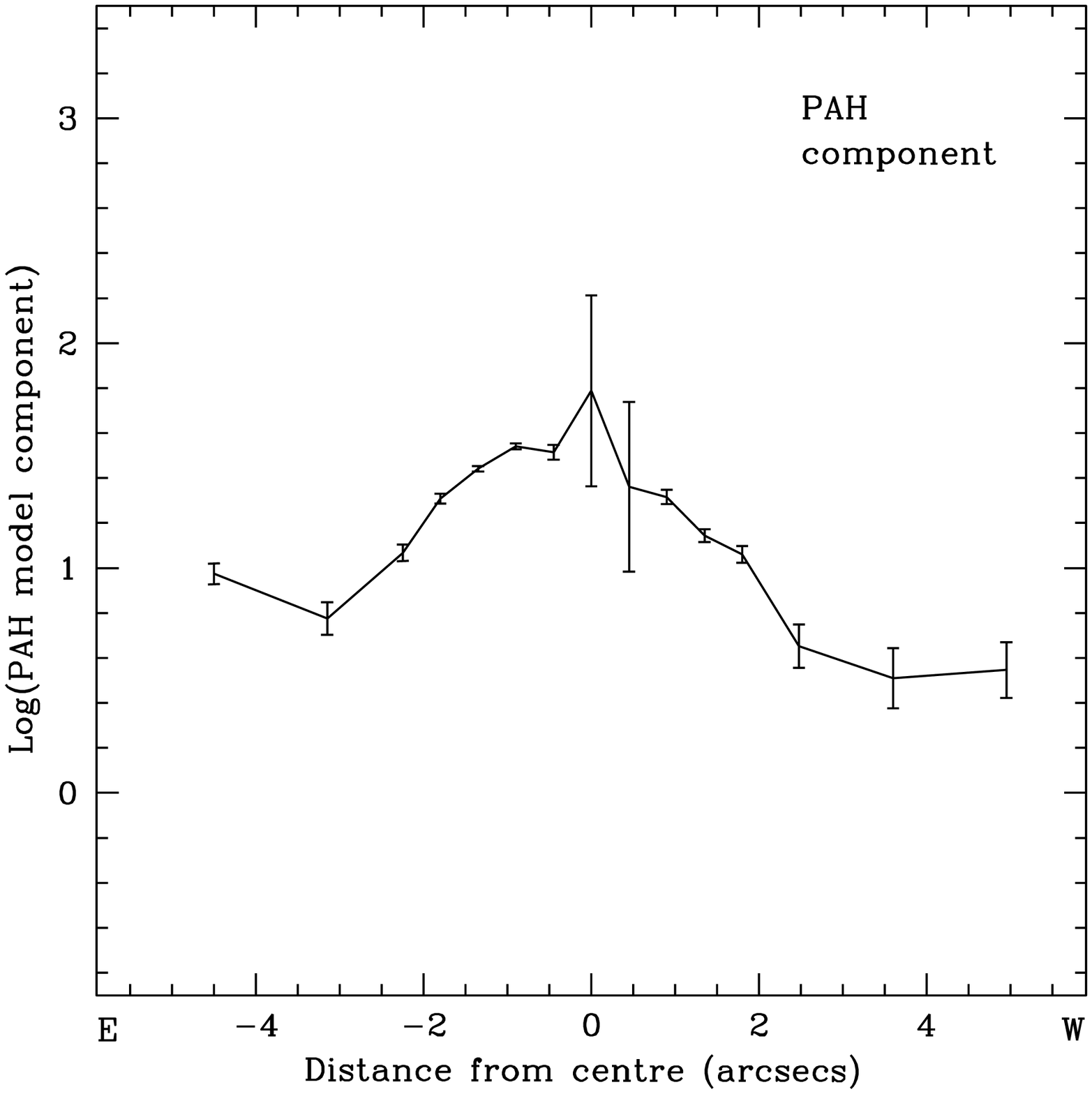}
\includegraphics[width=8.4cm]{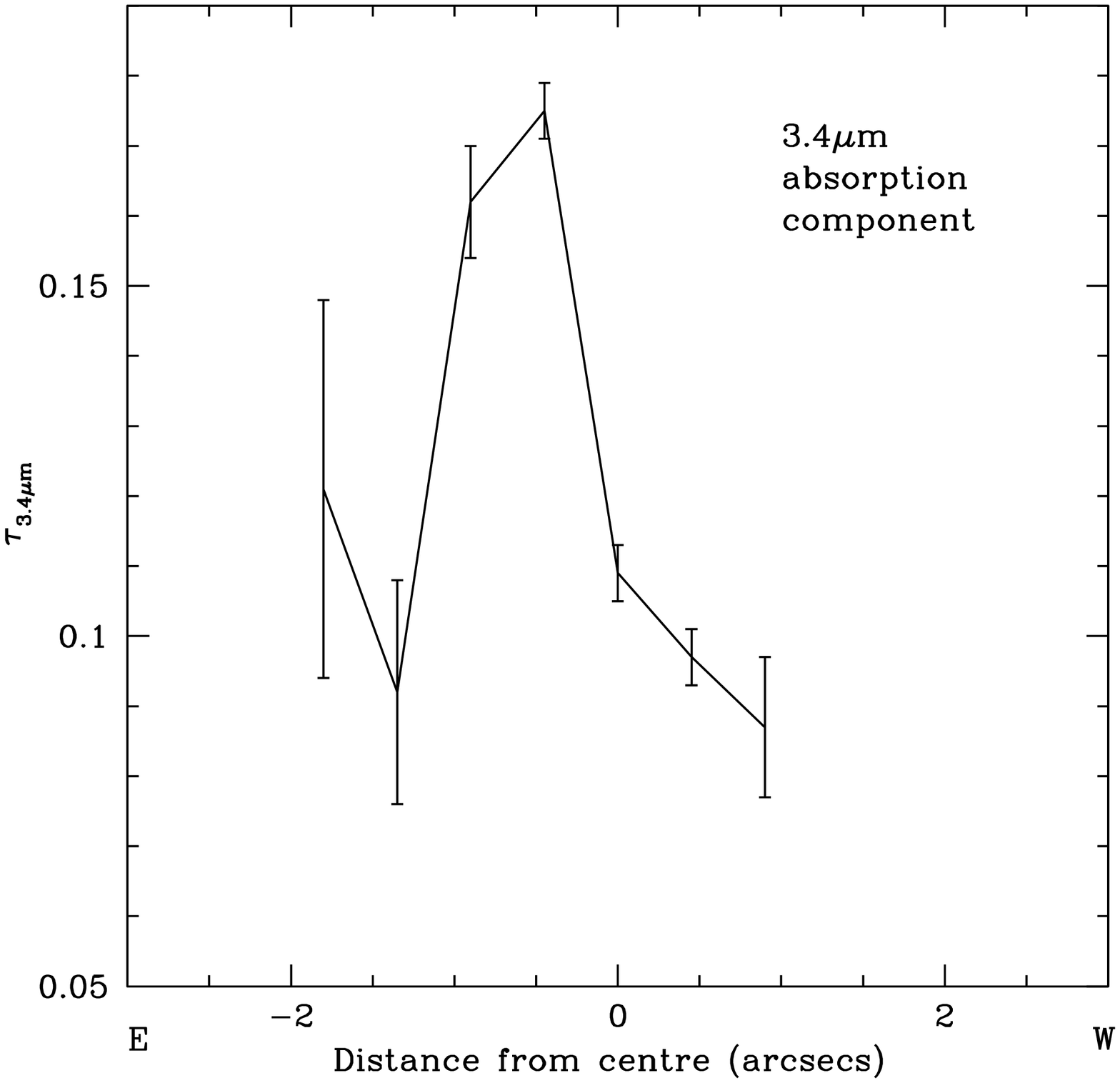}
\caption{The spatial variation of the 3.3-$\mu$m PAH emission flux (left) and the 3.4-$\mu$m hydrocarbon absorption (right) from the model fits. Errors are estimates from the model fitting process and do not fully reflect any systematic effects in the data. The PAH emission measured in the very centre is a small component against a large continuum and the uncertainties reflect this. Note that the spatial scale is smaller for the right-hand plot than in previous plots because the 3.4-$\mu$m absorption is undetected beyond 2 arcsec to the east and 1 arcsec to the west (see text).}
\label{fig:model_pahopdep}
\end{center}
\end{figure*}

Taking the peak flux to be coincident with the core, the core spectrum shows clear evidence of the 3.4-$\mu$m aliphatic hydrocarbon absorption. There the optical depth of the absorption is $\tau_{{3.4\mu}m}$ = 0.109~$\pm$0.004. This corresponds to $A_{v}$=15-25 mag using the Galactic dust extinction relation \citep{Pendleton94}. This is comparable to the minimum 30 mag extinction derived from the silicate absorption \citep{Roche06}. The right-hand figure of Fig. \ref{fig:model_pahopdep} shows how the optical depth varies with spatial position from the centre. The absorption rises to the east of the core, peaking at $\sim$0.5~-~0.9 arcsec, before decreasing beyond that. The absorption to the west appears to decrease with distance from the core and is undetected beyond 1 arcsec. These detection limits are due to both the decreasing absorption optical depth and the decreasing continuum contribution from the warm dust component against which the absorption can be detected, while the relative contribution from the diffuse hot component increases. The resulting spatial profile is consistent with there being a thick disc/torus shaped structure inclined at an angle similar to that of the galaxy. A structure of this type is also consistent with the silicate dust distribution, as suggested in \citet{Roche06}.

\begin{table}
\begin{center}
\begin{tabular}[tbp]{llll}
\hline
Distance from & $\tau_{{3.4\mu}m}$ &$\tau_{{9.7\mu}m}$ & $\tau_{{3.4\mu}m}$ \\
centre (arcsec) &&& / $\tau_{{9.7\mu}m}$ \\
\hline
\hline
-1.80 (E) & 0.121 & 2.16 & 0.056$\pm$0.0140\\
-1.35 & 0.092 & 2.64 & 0.035$\pm$0.0066\\
-0.90 & 0.162 & 2.45 & 0.066$\pm$0.0063\\
-0.45 & 0.175 & 2.04 & 0.086$\pm$0.0087\\
0.0 & 0.109 & 1.98 & 0.055$\pm$0.0059\\
0.45 & 0.097 & 1.91 & 0.051$\pm$0.0057\\
0.90 (W) & 0.087 & 1.28 & 0.068$\pm$0.0130\\
\hline
\end{tabular}
\caption{The spatial variation of the $\tau_{{3.4\mu}m}$/$\tau_{{9.7\mu}m}$ ratio in Circinus. $\tau_{{9.7\mu}m}$ values interpolated from \citet{Roche06}}
\label{tab:opdep}
\end{center}
\end{table}

Table \ref{tab:opdep} shows the spatial variation of the $\tau_{{3.4\mu}m}$/$\tau_{{9.7\mu}m}$ ratio, with the peak value at $\tau_{{3.4\mu}m}$/$\tau_{{9.7\mu}m}$=0.055. This is close to the Milky Way GC diffuse ISM ratio, which is 0.05-0.06 (using $\tau_{{3.4\mu}m}\sim$0.2 from \citet{Pendleton94} and $\tau_{{9.7\mu}m}\sim$3.6 from \citet{Roche85}). The Circinus ratio is also a relatively low value compared to observations of other nearby galaxies. \citet{Imanishi00} showed that the ratio of the absorption depths is spread from 0.02-0.23 between galaxies in his sample, with galaxies viewed at higher inclinations exhibiting lower ratios than those viewed face-on. Circinus is consistent with this finding, being at an inclination of 65$^{\circ}$ \citep{Freeman77}.

Whilst the optical depths of the silicate and 3.4-$\mu$m absorptions change by a factor of 2, the $\tau_{{3.4\mu}m}$/$\tau_{{9.7\mu}m}$ ratio remains approximately constant over a spatial scale of $\sim$60pc. This indicates that the aliphatic hydrocarbons and silicates are in one well-mixed dust population, with no evidence for any variation near the Seyfert nucleus.

\subsection{3.3$\mu$m PAH emission}
\label{sec:PAH}

The PAH component from the model fits (Fig. \ref{fig:model_pahopdep}) shows an emission plateau or gentle decline away from the nucleus, almost symmetric on each side. The strong continuum emission at the nucleus precludes a definitive detection of the 3.3-$\mu$m emission band in the core.
Its spatial dependence is similar to that of the hot blackbody component, suggesting that diffuse hot dust may contribute to both the continuum and the PAH emission. The profile is broadly similar to the 11.3-$\mu$m PAH emission, in both east/west and north/south alignments, from \citet{Roche06}, and very different from the distributions of the nuclear warm dust component. This clearly suggests a geometry different to the inner dust populations. The PAH emission region is not affected by the differential extinction of the aliphatic hydrocarbons and we conclude that it is therefore outside the primary absorption feature region. The finding corroborates that found in Circinus at 11.3-$\mu$m \citep{Roche06} and is consistent with the destruction of PAHs in high excitation regions \citep{Aitken85}. The spatial emission profile suggests that the PAH emission is excited primarily by diffuse emission.

\subsection{Line emission}

\begin{figure}
\begin{center}
\includegraphics[width=8.4cm]{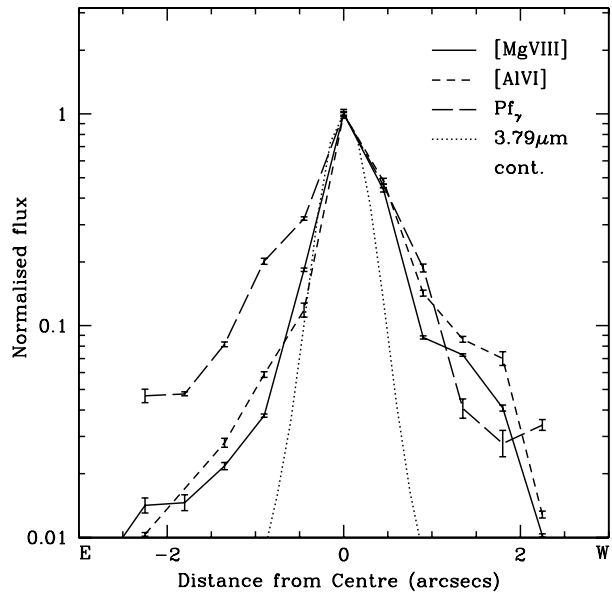}
\caption{Integrated fluxes of 3 emission lines and the PSF/continuum from the stellar calibrator. The fluxes have been normalized to allow comparison of the spatial profiles of the lines. The high-excitation forbidden lines are asymmetric, suggesting increased extinction to the east as discussed in the text. The {H\,\sc i} 8-5 line (Pf$_{\gamma}$) has a symmetrical profile yet is centrally peaked suggesting that the {H\,\sc ii} emission arises from a more diffuse component.}
\label{fig:All_lines}
\end{center}
\end{figure}

Line emission is detected to about two arcsec on each side of the nucleus, detections at the extremes partly dependent on the intrinsic flux of the line. Five lines have been firmly identified; two forbidden lines, [{Mg\,\sc viii}] at 3.03-$\mu$m and [{Al\,\sc vi}] at 3.66-$\mu$m, and three hydrogen Pfund-series lines, {H\,\sc i} 10-5 (Pf$_{\epsilon}$) at 3.04-$\mu$m, {H\,\sc i} 9-5 (Pf$_{\delta}$) at 3.30-$\mu$m and {H\,\sc i} 8-5 (Pf$_{\gamma}$) at 3.74-$\mu$m. The [{Al\,\sc viii}] line at 3.70-$\mu$m may also be present, but is coincident with a {Ca\,\sc i} absorption in the standard star and thus cannot be measured reliably. The spatially-extended emission of the most prominent lines is shown in Fig. \ref{fig:All_lines}.

\citet{Roche06} show that the emission from the 12.81-$\mu$m [{Ne\,\sc ii}] line and the 10.52-$\mu$m [{S\,\sc iv}] line is slightly stronger to the west of the centre than the east. The [{Si \,\sc vi}] (1.96-$\mu$m) and [{Al \,\sc ix}] (2.04-$\mu$m) lines have also been imaged by \citet{Maiolino98} and show spatially-extended emission, which in their fig. 4 shows greater emission to the west of the continuum peak. Similarly from this work, the forbidden lines [{Mg\,\sc viii}] and [{Al\,\sc vi}] show higher intensities to the west. This is interpreted as a result of increased extinction to the east, probably due to an extended dusty disk-like structure obscuring underlying symmetric emission regions. The flux ratios of the [{Mg\,\sc viii}], [{Al\,\sc vi}], [{S\,\sc iv}] and [{Ne\,\sc ii}] lines on the east and west sides of the nucleus, are approximately in accord with the wavelength dependence of extinction in the IR \citep{Lutz99}, consistent with the suggestion that IR extinction is largely responsible for the observed east/west asymmetry.

The Pfund series H lines show a different distribution to those of the forbidden lines. The strongest detection, Pf$_{\gamma}$, shows a centrally-peaked but roughly symmetric spatial distribution about the nucleus. The only comparable line from the literature is the Br$_{\gamma}$ line at 2.17-$\mu$m, imaged in Circinus by \citet{Maiolino98}. This too shows a symmetric distribution about the peak with a spatially extended profile, as seen in Fig. \ref{fig:All_lines}. This naturally suggests that there are contributions to the Pf$_{\gamma}$ flux from both a high excitation nuclear component and from diffuse emission largely outside the cold dust extinction region shown by the 3.4-$\mu$m absorption. Hydrogen in this region is presumably primarily ionized by flux from the circumnuclear star formation. 

\section{Discussion}

\begin{figure}
\begin{center}
\includegraphics[width=8.4cm]{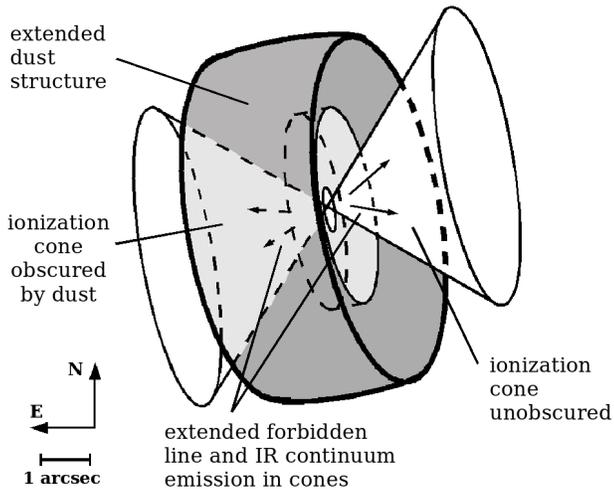}
\caption{Sketch of the proposed extended dust structure for Circinus from 3 and 10-$\mu$m observations. The inner IR continuum emission is from dust heated by radiation from the core, while the extended east-west continuum emission is from dust heated in the ionization cones. An inclined dusty thick structure produces the differential east-west extinction (see text).}
\label{fig:sketch}
\end{center}
\end{figure}

The origin of the extended T$\sim$500~K dust emission and extended coronal line emission presented above is not clearly understood. Emission at $\sim$100$^\circ$ PA extending to the east and west for $>$1 arcsec is clearly present in the 10-$\mu$m and 20-$\mu$m bands \citep{Packham05}. A similar structure is evident in the {[Si\,\sc VII]} image from \citet{Prieto04} while there is a suggestion of it in their $M'$ 4.8-$\mu$m band image. This extended emission is not visible in the $L$-band image of \citet{Prieto04}, presumably because it is swamped by the stronger diffuse component, which is separated out in our spectra by the model fitting. The extended dust emission is coincident with the coronal line emitting region and therefore a reasonable explanation is that dust is heated by trapped line emission in the ionization cones. The emitting dust must be refractory and able to survive in this harsh environment, and provides a background source against which the 3.4-$\mu$m absorption can be measured. The symmetry of the 8.8-$\mu$m image of \citet{Packham05} to the east and west suggests that the intrinsic $L$-band continuum dust emission is also symmetric across the nucleus. Variations in the emission line intensities and the 3.4-$\mu$m hydrocarbon absorption depth are then primarily due to variations in extinction due to changes in the absorbing dust column along the line of sight. An illustration of such a configuration is shown in Fig. \ref{fig:sketch}.

The ratio of the hydrocarbon to silicate dust optical depths, $\tau_{{3.4\mu}m}$/$\tau_{{9.7\mu}m}$, is fairly constant across a projected distance of 60~pc in the nuclear region of Circinus. This compares to significant variations found between galaxies by \citet{Imanishi00}. Imanishi suggested that a possible explanation for these variations could be that the 9.7-$\mu$m silicate dust column probes the line of sight to dust emission at T$\sim$300~K, while the 3.4-$\mu$m hydrocarbon band probes to dust emission at T$\sim$ 1000~K, and that temperature gradients in the circumnuclear dust could then produce variations in the $\tau_{{3.4\mu}m}$/$\tau_{{9.7\mu}m}$ ratio. While, the observations presented here are not sensitive to temperature gradients within the central 10pc, we see no evidence for temperature gradients in the dust emission beyond that and no evidence of variations in $\tau_{{3.4\mu}m}$/$\tau_{{9.7\mu}m}$. It seems that the paths to the 3$~\mu$m and 10$~\mu$m emitting regions in Circinus are similar, and that the hydrocarbon and silicate components of the dust are well mixed, with a ratio similar to that found towards the Galactic centre in the Milky Way. It is striking that any variations within the $\tau_{{3.4\mu}m}$/$\tau_{{9.7\mu}m}$ ratio in Circinus are much smaller than the variations between galaxies \citep{Imanishi00}

The 3.3-$\mu$m PAH emission band has a very different and much broader spatial distribution to the continuum emission. It is similar to that of the 11.3-$\mu$m feature \citep{Roche06} and to the diffuse galactic emission, suggesting that it is excited primarily by a diffuse radiation field rather than the AGN, and in accord with the expectation that the acrriers are destroyed by the hard photons around the nucleus \citep{Aitken85}.

\section{Conclusions}

Long slit spectra of the Circinus galaxy between 2.9 and 3.9~$\mu$m at a spatial resolution of $\sim$0.5 arcsec are presented. The continuum emission from the nuclear region is separated into two components, a compact core with dust emission at T$\sim$540~K and a diffuse component, represented by a T=1800~K blackbody. The flux from the compact core and the extended east-west structure dominates the emission within the central $\sim$3 arcsec. The 3.4-$\mu$m aliphatic hydrocarbon absorption has been measured towards this emission structure. The optical depth is $\tau_{{3.4\mu}m}\sim$~0.1 towards the nucleus, but changes by a factor of 2 in the central 60~pc, decreasing to the west but increasing to the east. Similar variations have been found in the depth of the 9.7-$\mu$m silicate absorption \citep{Roche06}, while the asymmetry in the [{Mg\,\sc viii}] and [{Al\,\sc vi}] emission line intensities on either side of the nucleus also suggests significant differential extinction. All of these observations are consistent with a coronal line region containing warm emitting dust and extending east and west from the nucleus. The nucleus is obscured with A${\rm _V} \sim$ 25~mag, which increases to $\sim 40$ mag at 1 arcsec to the east and decreases to the west. This may indicate an inclined dusty disc-like structure, with the east side inclined toward us and lying behind the coronal line region on the west side.   Large scale dusty disks appear to be a common feature in the few AGN that have been probed spectroscopically at subarcsecond  resolution (e.g.  \citet{Mason06}, \citet{Roche06}, \citet{Young07}), and adaptive optics correction would permit higher resolution measurements of the dust properties in the L-band.

The ratio of $\tau_{{3.4\mu}m}$/$\tau_{{9.7\mu}m}\sim$~0.06 for Circinus is comparable to the Galactic value \citep{Pendleton94}. It remains approximately constant within the region for which both absorptions are detected, from $\sim$40 pc east to 20 pc west ($\sim$90 and 50 pc projected distance on a 65$^{\circ}$ disc). This indicates that the two dust populations are well-mixed within the central region and neither is preferentially destroyed or suppressed. The PAH population, however, has a flatter spatial emission profile across the inner $\pm$2 arcsec, indicating PAHs are not primarily excited by the nuclear flux or are absent in the inner region. The PAH distribution more closely resembles the diffuse galactic emission and is likely to be related to the circumnuclear star-formation regions \citep{Marconi94}.

\section*{Acknowledgments}

This work is based on observations obtained at the Gemini Observatory, which is operated by the
Association of Universities for Research in Astronomy, Inc., under a cooperative agreement
with the NSF on behalf of the Gemini partnership: the National Science Foundation (United
States), the Science and Technology Facilities Council (United Kingdom), the
National Research Council (Canada), CONICYT (Chile), the Australian Research Council
(Australia), Minist\'{e}rio da Ci\^{e}ncia e Tecnologia (Brazil) and SECYT (Argentina). This work has made use of the NASA/IPAC Extragalactic Database (NED) which is operated by the Jet Propulsion Laboratory, California Institute of Technology, under contract with the National Aeronautics and Space Administration. We thank the Gemini Observatory and particularly the staff at Gemini-S for assistance in collecting these data. M. C. thanks the U. K. Science and Technology Facilities Council for the support of a studentship. We thank the referee for helpful suggestions.

\label{lastpage}

\end{document}